\documentclass[twoside,twocolumn,9pt]{article}
\usepackage{extsizes}
\usepackage[super,sort&compress,comma]{natbib}
\usepackage[version=3]{mhchem}
\usepackage[left=1.5cm, right=1.5cm, top=1.785cm, bottom=2.0cm]{geometry}
\usepackage{balance}
\usepackage{mathptmx}
\usepackage{sectsty}
\usepackage{graphicx} 
\usepackage{lastpage}
\usepackage[format=plain,justification=justified,singlelinecheck=false,font={stretch=1.125,small,sf},labelfont=bf,labelsep=space]{caption}
\usepackage{float}
\usepackage{fancyhdr}
\usepackage{fnpos}
\usepackage[english]{babel}
\addto{\captionsenglish}{%
  
}
\usepackage{array}
\usepackage{charter}
\usepackage[T1]{fontenc}
\usepackage[usenames,dvipsnames]{xcolor}
\usepackage{setspace}
\usepackage[compact]{titlesec}
\usepackage{hyperref}

\definecolor{cream}{RGB}{222,217,201}

\begin{document}

\pagestyle{fancy}
\thispagestyle{plain}
\fancypagestyle{plain}{
%%%HEADER%%%
\renewcommand{\headrulewidth}{0pt}
}
%%%END OF HEADER%%%

%%%PAGE SETUP - Please do not change any commands within this section%%%
%\newcommand{\NEW}[1]{{\color{red} #1}}
%\newcommand{\OLD}[1]{\sout{#1}}
%\newcommand{\KED}[1]{{\color{magenta} {\bf #1}}}

\makeFNbottom
\makeatletter
\renewcommand\LARGE{\@setfontsize\LARGE{15pt}{17}}
\renewcommand\Large{\@setfontsize\Large{12pt}{14}}
\renewcommand\large{\@setfontsize\large{10pt}{12}}
\renewcommand\footnotesize{\@setfontsize\footnotesize{7pt}{10}}
\makeatother
\renewcommand{\thefootnote}{\fnsymbol{footnote}}
\renewcommand\footnoterule{\vspace*{1pt}% 
\color{cream}\hrule width 3.5in height 0.4pt \color{black}\vspace*{5pt}} 
\setcounter{secnumdepth}{5}

\makeatletter 
\renewcommand\@biblabel[1]{#1}            
\renewcommand\@makefntext[1]% 
{\noindent\makebox[0pt][r]{\@thefnmark\,}#1}
\makeatother 
\renewcommand{\figurename}{\small{Fig.}~}
\sectionfont{\rmfamily\Large}
\subsectionfont{\normalsize}
\subsubsectionfont{\bf}
\setstretch{1.125} %In particular, please do not alter this line.
\setlength{\skip\footins}{0.8cm}
\setlength{\footnotesep}{0.25cm}
\setlength{\jot}{10pt}
\titlespacing*{\section}{0pt}{4pt}{4pt}
\titlespacing*{\subsection}{0pt}{15pt}{1pt}
%%%END OF PAGE SETUP%%%

%%%FOOTER%%%
\fancyfoot{}
\fancyfoot[LO,RE]{\vspace{-7.1pt}\includegraphics[height=9pt]{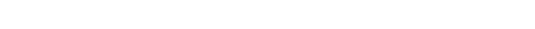}}
\fancyfoot[CO]{\vspace{-7.1pt}\hspace{13.2cm}\includegraphics{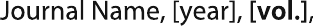}}
\fancyfoot[CE]{\vspace{-7.2pt}\hspace{-14.2cm}\includegraphics{head_foot/RF}}
\fancyfoot[RO]{\footnotesize{\sffamily{1--\pageref{LastPage} ~\textbar  \hspace{2pt}\thepage}}}
\fancyfoot[LE]{\footnotesize{\sffamily{\thepage~\textbar\hspace{3.45cm} 1--\pageref{LastPage}}}}
\fancyhead{}
\renewcommand{\headrulewidth}{0pt} 
\renewcommand{\footrulewidth}{0pt}
\setlength{\arrayrulewidth}{1pt}
\setlength{\columnsep}{6.5mm}
\setlength\bibsep{1pt}
%%%END OF FOOTER%%%

%%%FIGURE SETUP - please do not change any commands within this section%%%
\makeatletter 
\newlength{\figrulesep} 
\setlength{\figrulesep}{0.5\textfloatsep}

\makeatother
%%%END OF FIGURE SETUP%%%

%%%TITLE, AUTHORS AND ABSTRACT%%%
\twocolumn[
  \begin{@twocolumnfalse}
%{\includegraphics[height=30pt]{head_foot/SM}\hfill\raisebox{0pt}[0pt][0pt]{\includegraphics[height=55pt]{head_foot/RSC_LOGO_CMYK}}\\[1ex]
%\includegraphics[width=18.5cm]{head_foot/header_bar}}\par
\vspace{1em}
\rmfamily
\begin{tabular}{m{4.5cm} p{13.5cm} }

\includegraphics{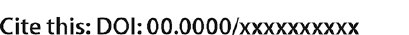} & \noindent\LARGE{\textbf{Controlling rheology via boundary conditions in dense granular flows$^\dag$}} \\%Article title goes here instead of the text "This is the title"
\vspace{0.3cm} & \vspace{0.3cm} \\

 & \noindent\large{Farnaz Fazelpour, and Karen E. Daniels\textit{$^{a}$}} \\%Author names go here instead of "Full name", etc.

\includegraphics{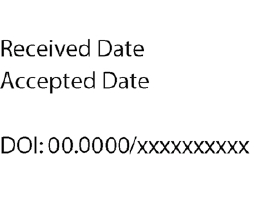} & \noindent\normalsize{
Boundary shape, particularly roughness, strongly controls the amount of wall slip in dense granular flows. In this paper, we aim to quantify and understand which aspects of a dense granular flow are controlled by the boundary condition, and to incorporate these observations into a cooperative nonlocal model characterizing slow granular flows. To examine the influence of boundary properties, we perform experiments on a quasi-2D annular shear cell with a rotating inner wall and a fixed outer wall; the later is selected from among 6 walls with various roughness, local concavity, and compliance. We find that we can successfully capture the full flow profile using a single set of empirically determined model parameters, with only the wall slip velocity set by direct observation. Through the use of photoelastic particles, we observe how the internal stresses fluctuate more for rougher boundaries, corresponding to lower wall slip, and connect this observation to the propagation of nonlocal effects originating at the wall.} \\

\end{tabular}

 \end{@twocolumnfalse} \vspace{0.6cm}

  ]
%%%END OF TITLE, AUTHORS AND ABSTRACT%%%

%%%FONT SETUP - please do not change any commands within this section
\renewcommand*\rmdefault{bch}\normalfont\upshape
\rmfamily
\section*{}
\vspace{-1cm}

%%%FOOTNOTES%%%

\footnotetext{\textit{$^{a}$~Department of Physics, North Carolina State University, Raleigh, NC, USA.}}

%Please use \dag to cite the ESI in the main text of the article.
%If you article does not have ESI please remove the the \dag symbol from the title and the footnotetext below.
\footnotetext{\dag~Electronic Supplementary Information (ESI) available. See DOI: }

%%%END OF FOOTNOTES%%%

%%%MAIN TEXT%%%%
%The main text of the article\cite{Mena2000} should appear here.

\section{Introduction}

Understanding the no-slip boundary condition has been instrumental to the development of the field of fluid dynamics\cite{lauga_microfluidics_2007}. In granular flows, predicting the amount of slip at a boundary has been challenging, as neither the no-slip condition nor Coulomb friction are valid \cite{artoni_effective_2009}. Furthermore, boundary roughness can affect not only the amount of slip near the wall, but also the bulk behavior of granular material.\cite{silbert_boundary_2002} Boundary conditions have been found to impact fluidization and create new plastic events in emulsions\cite{mansard_boundary_2014}, and such nonlocal effects are likely present in granular materials as well. Because stress propagates through heterogeneous force chains, the boundary conditions effects can influence distant neighbor grains, with either small or large effects depending on the particular pathways of the force chains. Detrimental flow behaviors such as intermittency\cite{silbert_temporally_2005}, creeping/stagnant zones\cite{reddy_evidence_2011,amon_experimental_2013,deshpande_perpetual_2021}, and clogging\cite{zuriguel_invited_2014,mort_dense_2015,tang_how_2011} arise from a complex interaction between the flow-forcing and the boundary conditions.

Traditionally, granular materials have been described using local rheology, which characterizes granular flows using two dimensionless, locally-determined variables. The first, inertial number $I$, describes the flow rate\cite{forterre2008flows}:
\begin{equation}
I \equiv \frac{\dot{\gamma} d}{\sqrt{P/\rho}}
\label{eq:I}
\end{equation}
This quantity arises from examining the ratio between two time scales: a microscopic time scale $T=d/\sqrt{P/\rho}$ which is the time to squeeze a particle into a hole of the same size ($d$ is the particle diameter, $P$ is the pressure, $\rho$ is the particle density) and a macroscopic time scale $1/\dot{\gamma}$ which is the deformation time under shear rate $\dot{\gamma}$. The second dimensionless variable is the stress ratio $\mu$, the ratio between the shear stress $\tau$ and pressure $P$:
\begin{equation}
\mu \equiv \frac{\tau}{P}.
\label{eq:mu}
\end{equation}
In a local rheology, $\mu$ plays the role of an effective friction\cite{midi2004dense}, where there is no flow below a yield criterion $\mu_{s}$, \cite{da2005rheophysics}. For $\mu > \mu_s$, the inertial number $I$ is proportional to the excess $\mu$, with a constant $b$ controlling the steepness of this relationship. The relationship $I(\mu)$ can therefore be written using the Heaviside function $H$:
\begin {equation}
I(\mu)=\frac{\mu-\mu_s \, }{b} H(\mu-\mu_s).
\label{eq:k1}
\end {equation}

Even though local rheology can explain many features of granular flows\cite{forterre2008flows}, it cannot capture some experimental observations \cite{koval2009annular,midi2004dense,Cheng2006,nichol2010flow,reddy_evidence_2011}, particularly where fluctuations or vibrations play a role. To address these shortcomings, several nonlocal rheology models have been developed\cite{Bouzid2013,bouzid2015non,Kamrin2012,henann2013predictive,zhang2016,dsouza_non-local_2020}. The nonlocal rheology model we are focusing on in this paper is the cooperative model developed by \citet{Kamrin2012}, which extends a Bagnold-type granular flow via the diffusion of fluidity. We have previously observed that this model can successfully capture granular flows in a 2D annular rheometer across various packing fractions and shearing rates,\cite{tang2018nonlocal} as well as for a variety of particle stiffnesses and shapes.\cite{fazelpour_effect_2022} In both cases, a single set of nonlocal parameters can be assigned to a particular set of particles. However, in order for a nonlocal model to be a successful predictive tool, it will be necessary to make flow measurements for a set of particles in one geometry, and then use them to provide predictions for flows in other geometries. Currently, good fits of the model to the data depend on a priori knowledge of the amount of wall slip, limiting its predictive power.

\begin{figure} 
	\centering 
	\includegraphics[width=1.1\linewidth]{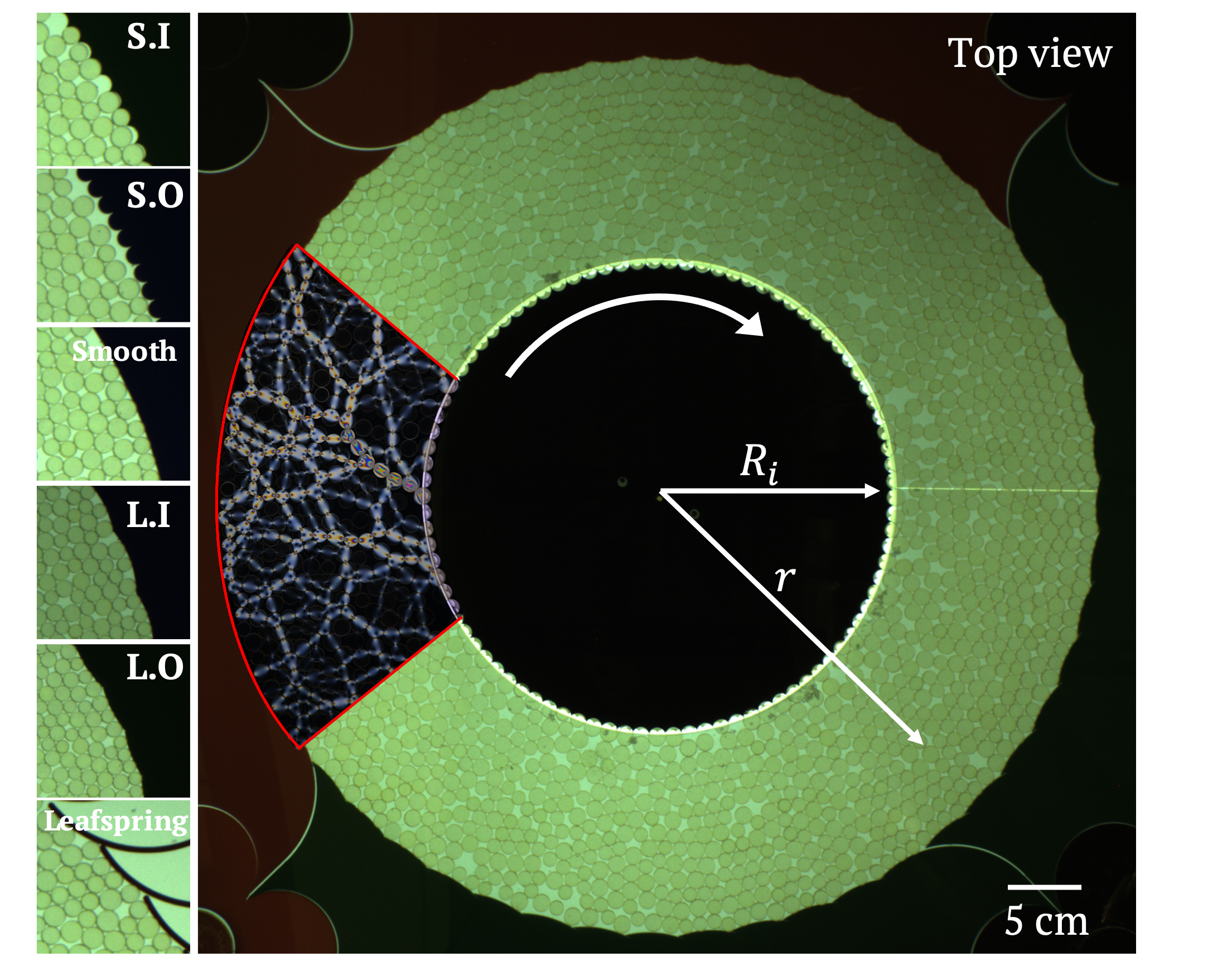} 
	\caption{Right: Top view of 2D annular shear cell. The inner wall ($R_{i}=15$~cm) rotates at constant speed and in a clock-wise direction. The coordinate $r$ measures the distance from the center of inner wheel. Inside the red box: Sample image of photoelastic particles sandwiched between two polarizers to visualize the force chains. Left: Six different outer walls used in the experiment with various roughness/compliance, labelled with the names used in the figure legends. 
		}
	\label{fig:apparatus} 
\end{figure}

In this paper, we aim to understand which aspects of a dense granular flow are controlled by the roughness of the boundaries, and incorporate those observations into the cooperative model. We conduct experiments with six different wall roughnesses, local concavity, and compliance using the same particles (see Fig.~\ref{fig:apparatus}), each resulting in a different amount of wall slip. We quantify how the cooperative model responds to various boundary properties and characterize flow features as a result of various boundaries. While the cooperative model is again successful in describing granular flow for various boundary properties, it is still necessary to measure the amount of wall slip for each experiment. Motivated by \citet{thomas_photoelastic_2019}, we aim to resolve this issue by quantifying the amount of wall slip as a function of both the inter-particle force fluctuations and the particular boundary roughness which affects them. We reveal that there is an excess force fluctuation due to wall roughness and compliance which allows us to qualitatively predict the wall slip for each boundary and find boundary properties to be a source of nonlocal effects.

\subsection{The cooperative model}\label{subsection}

The cooperative model is a nonlocal rheology model developed by \citet{Kamrin2012} to capture granular flows at intermediate to creeping rates\cite{Kamrin2012,henann2013predictive,zhang2016}. It is a continuum model developed by extending a local Bagnold-type granular flow to include nonlocal effects considerations. Under Bagnold scaling, there is a linear relationship between shear stress and shear rate. This gives rise to the definition of fluidity:
\begin{equation}
g \equiv \frac{\dot{\gamma}}{\mu}
\label{eq:g}
\end{equation}
where $\dot{\gamma}$ is the shear rate and $\mu$ is the stress ratio defined in Eq.~\ref{eq:mu}.
Taking only local information into account, the local fluidity $g_\mathrm{loc}$ can be obtained using locally-determined variables (substituting Eq.~\ref{eq:I} and Eq.~\ref{eq:k1} into Eq.~\ref{eq:g}):
\begin {equation}
g_\mathrm{loc}(\mu,P) = \frac{\mu-\mu_s}{b \mu T}  H(\mu-\mu_s)
\label{eq:k2}
\end {equation}
However, in granular materials, particle rearrangements in one part of a flow can trigger (or suppress) rearrangements elsewhere, by moving a particular contact closer to, or further from, failure. Therefore, the flow-resistance should be a function of both the local shear rate and these nonlocal events. The cooperative model takes into account these nonlocal events by including a Laplacian term to account for the diffusion of nonlocal effects: 
\begin {equation}
\nabla^2 g = \frac 1 {\xi ^2}(g-g_\mathrm{loc}).
\label{eq:k3}
\end {equation}
The diffusion length scale $\xi$ describes how far away these rearrangements can influence the rest of the material. Thus, the nonlocal model introduces a new characteristic mesoscopic lengthscale between the particle (micro) and bulk (macro) scales. The length scale $\xi$ is measured in units of the particle diameter $d$ and is defined as:
\begin {equation}
\frac{\xi}{d} = A\sqrt{\frac{1}{|\mu-\mu_s|}}.
\label{eq:k4}
\end {equation}
where the nonlocal parameter $A$ is an experimentally-determined constant for a particular set of particles \cite{fazelpour_effect_2022} and indicates the strength of their nonlocal effects. $\xi$ is symmetric and most sensitive near the yield criterion $\mu_{s}$. 
 
 The cooperative model has been validated in both 2D and 3D simulations \cite{kamrin2014effect,Kamrin2012,henann2013predictive} and in 2D experiments (annular Couette)\cite{tang2018nonlocal, fazelpour_effect_2022}. While $b$ need only be a quantity near unity, the local and nonlocal parameters $\mu_{s},A$ are constants and a function of both material properties and particle shape\cite{fazelpour_effect_2022}; they need to be determined separately, for each new set of particles. Once these parameters are known, the cooperative model can model granular flows by solving Eq.~\ref{eq:k3} with the boundary condition (wall slip) determined empirically.
 
%===============================================================
\section{Method}

\subsection{Apparatus \label{sec:apparatus}}

\begin{table*}
	\centering
	\begin{tabular}{|c||c|c|c|c|c|c|c|c|}
		\hline    
		wall name &S.I. &S.O. &smooth &L.I.&L.O.&leafspring  \\ \hline
		shape &small innies &small outies &smooth &large innies &large outies & large outies\\ \hline
		rigidity &rigid &rigid &rigid&rigid &rigid & compliant\\ \hline
		$v(R_i)$ [$d$/s]  &1.1 &1.1 &1.1 &1.1&1.1 &1.1  \\ \hline
		\# of particles       &1784 &1740 &1715 &1747  &1724 &1724  \\ \hline
		$\phi$ & 0.65  & 0.65& 0.65& 0.65&  0.65 & 0.65 \\ \hline    
	\end{tabular}
	\caption{{\bf Summary of the datasets.} The first row contains the name of each wall shape used in the figure legends, followed by two rows with containing a more detailed description of the length scale, curvature, and compliance. The final three rows give the inner wall rotation speed $v(R_i)$ (held constant), the number of particles, and the target packing density (held constant).}
	\label{tbl:packingdensity}
\end{table*}

Our apparatus is a quasi-2D annular shear cell, shown in Fig.~\ref{fig:apparatus}. The central disk provides a driven inner wall with a radius $R_{i}=15$~cm, attached to a motor (Parker Compumotor BE231FJ-NLCN with a PV90FB 50:1 gearbox). The motor rotates the inner wall with a constant speed ($v = 1.1$~cm/s for all experiments presented here). The inner wall is roughened with holes matching the width of a single particle, providing the shear force that generates the observed flow. The annular geometry allows for continuous shearing; all the data reported in this paper are taken at steady state, after the flow is fully developed.

The stationary outer wall is located at radius $R_{o}=28$~cm, measured from the center of the inner wall. The outer walls are each laser-cut with a specific roughness, curvature, and compliance, providing the six different patterns shown in Fig.~\ref{fig:apparatus}. Of these, one is rigid and smooth, one is composed of compliant leafsprings, two more rigid wall patterns have the same shape taken from the compliant leafsprings (both denoted L = long for their length scale), and two rigid walls have the same shape taken from the diameter of the particles (denoted S = short). The curvature of the four solid walls is denoted as O = outies and I = innies. 
The 52 leafsprings, each of which compresses approximately linearly under stress, are the same walls used in \citet{tang2018nonlocal,fazelpour_effect_2022}. 
In all the datasets, the inner wall has an S.I. pattern, with only the outer walls changing. 

The particles used in all six datasets are photoelastic disks made of Vishay PhotoStress material PSM-4 (bulk modulus $E=4$~MPa and density $\rho=1.06$~g/cm$^3$). These are the same particles used in \citet{owens_acoustic_2013,owens_sound_2011,fazelpour_effect_2022}. The mixture has an equal population of bidisperse circles with diameter $d_S = 0.9$~cm and $d_L = 1.1$~cm and thickness $6.35$~mm. The optical properties of this photoelastic material allows for visualizing the internal stress between particles throughout the material, as shown inside Fig.~\ref{fig:apparatus}. 
Our data is collected using a color-separation technique \cite{kollmer_photo-elastic_nodate,fazelpour_effect_2022} in which monochromatic, unpolarized red light provides the images used for particle-finding, and monochromatic, polarized green light provides the images used for measuring internal stresses.
We measure the vector force at each contact by solving an inverse problem using the observed fringe patterns within each particle. More details about this technique are available in \cite{abed_zadeh_enlightening_2019,kollmer_photo-elastic_nodate,liu_spongelike_2021,daniels_photoelastic_2017,fazelpour_effect_2021}. From these vector contact forces, we coarse-grain \cite{weinhart2013coarse, fazelpour_effect_2021} the dataset to provide radial profiles of the two independent moments of the stress tensor: shear stress $\tau(r)$ and pressure $P(r)$.

Before running each new set of walls, we adjust the number of particles to keep the packing density $\phi=0.65$ constant for all 6 wall shapes, each of which has a slightly different volume (area) even though they have the same $R_{o}$. To precisely measure the volume of each system, we reference each of the other walls to the smooth wall, for which the areas between the inner wall and outer wall is easiest to calculate. We photograph each of the other walls on top of the smooth wall; because they are cut from different color acrylic, we can use color separation and image processing to measure the difference in area and adjust the number of particles accordingly. The only exception is the leafspring walls: due to compliance of the springs, a constant packing density is not possible. Instead we use the same number of particles as for the large outies (L.O.) since these two walls have the same shape/curvature in the relaxed state.

\subsection{Speed and shear rate measurements\label{sec:speed}}

\begin{figure}
	\centering 
	\includegraphics[width=0.9\linewidth]{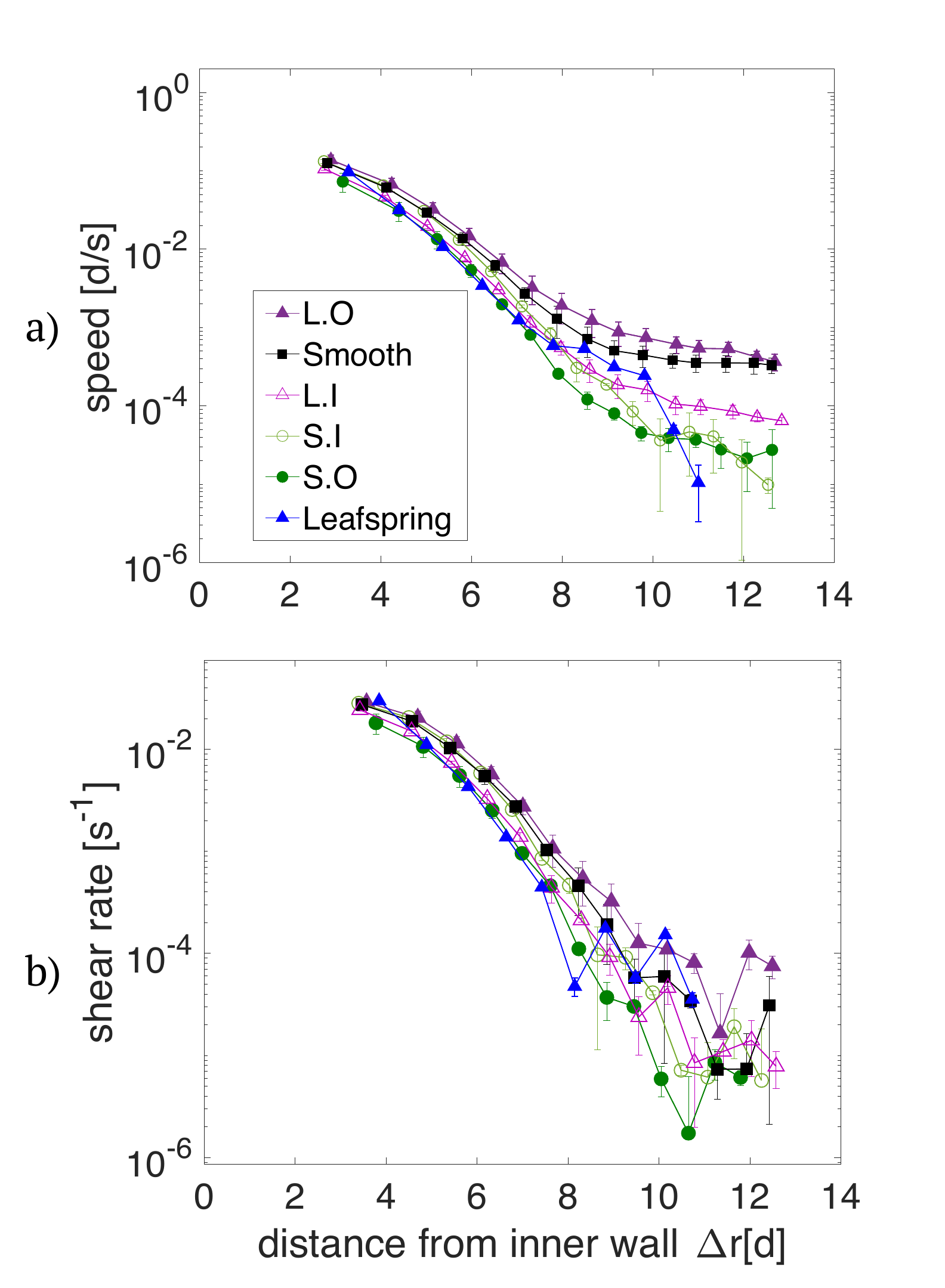} 
	\caption{(a) Tangential speed profile $v(\Delta r)$ and (b) shear rate $\dot{\gamma}(\Delta r)$, both on a logarithmic scale, as a function of distance from the inner wall $\Delta{r}= r-R_{i}$, for all six outer wall shapes. The logarithmic axes were chosen to highlight the flow behavior in the creeping (slow) region of the flow near the outer wall. Error bars are standard errors. The leafspring data are from \citet{fazelpour_effect_2022}.
	}
	\label{fig:speed} 
\end{figure}

To measure the flow profile, we first find the locations of the particles using the Matlab Hough transform package \cite{HT} and then implement the Blair-Dufresne particle-tracking algorithm \cite{DB} to obtain the trajectory of each particle. We calculate the average particle velocity within concentric rings of width $0.65d$ to obtain the tangential speed $v$ as a function of distance from inner wall $\Delta r$. From this binned data, we also calculate the velocity variance $\delta v^2$ for later use in granular temperature measurements; note that this is only one of two velocity components. Finally, we measure the shear rate profile using Fourier-derivatives of the speed profile as described in \citet{tang2018nonlocal}, according to ${\dot{\gamma}}(r) = \frac{1}{2} \left( \frac {\partial v}{\partial r}- \frac{v}{r} \right)$ in polar coordinates.

We analyze and average the speed and shear rate profiles for $10^4$ images for each wall shape. As shown in Fig.~\ref{fig:speed}a, the speed profiles near the inner wall (fastest flow) are quite similar for all 6 wall shapes. However, in the creeping regime close to the outer wall, we observe significant differences: the roughest walls (S.I. and S.O.) have much less wall slip than the smoothest walls (L.O. and smooth), with both curvature and length scale playing a role. Furthermore, the leafspring walls suppress slip, likely due to springs' deformation allowing for a force to be applied in the direction opposing the flow. In contrast, we observe that the shear rate profiles $\dot{\gamma}$ for the different wall shapes are more similar to each other (see Fig.~\ref{fig:speed}b) than are the speeds.

\subsection{Shear stress and pressure measurements\label{sec:shear}}

Measuring the stress field within a granular experiment has been a longstanding challenge. In most experimental studies, stress fields are measured at the boundaries and predicted throughout the material by considering physical and chemical interactions in the system. Here, we are able to employ photoelastic techniques to obtain stress fields throughout the material, as illustrated via the bright pattern of force chains in Fig.~\ref{fig:apparatus}. 

\begin{figure} 
\centering 
\includegraphics[width=0.9\linewidth]{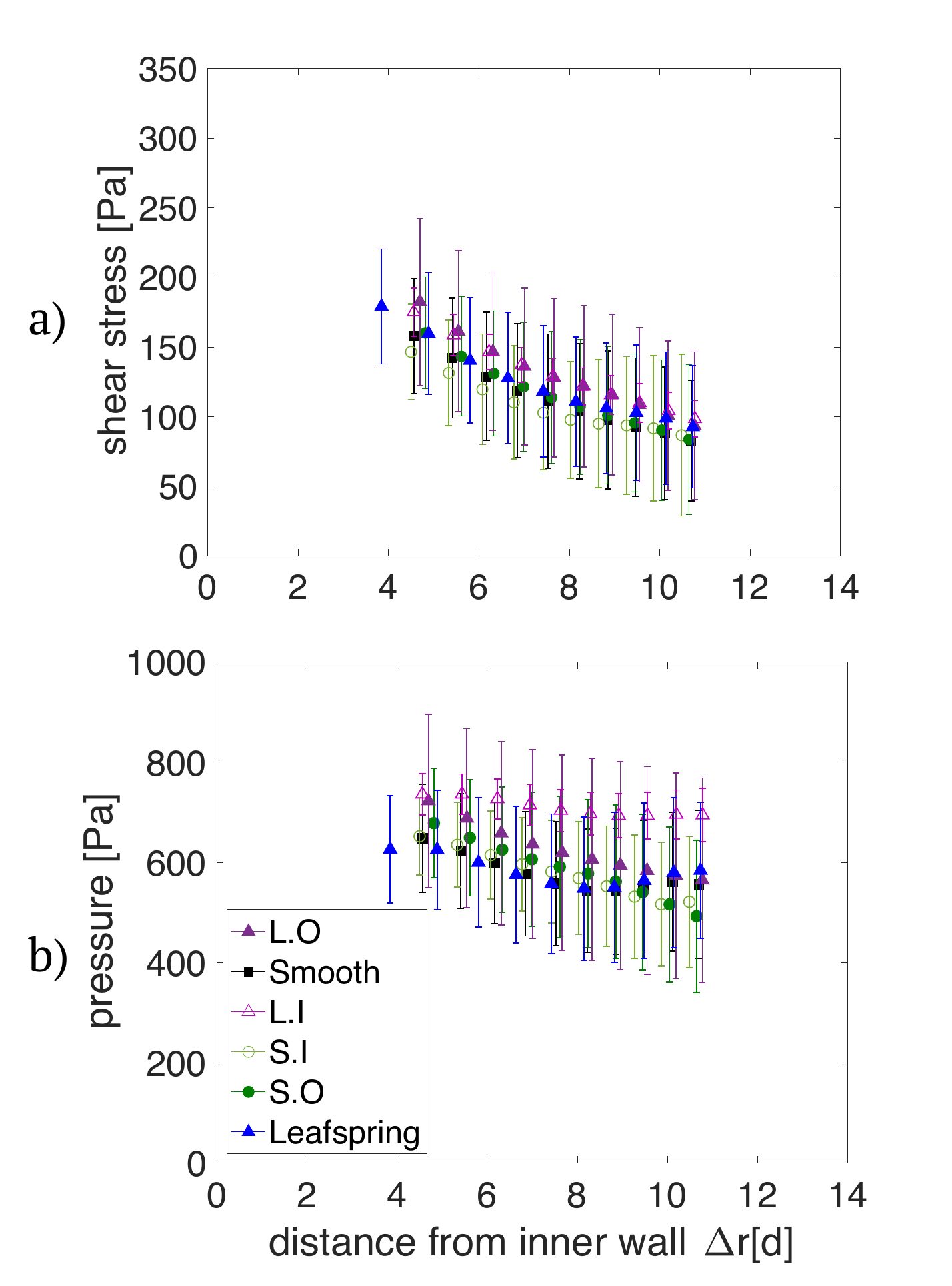} 
\caption{(a) Shear stress profiles $\tau(\Delta r)$ and (b) pressure profiles $P(\Delta r)$, both measured using photoelasticity, as a function of distance from inner wall $\Delta r$, for all 6 wall shapes. The error bars indicate spatial and temporal standard deviations. The leafspring data are from \citet{fazelpour_effect_2022}}
 \label{fig:stress} 
\end{figure}

Using color-separated images of the polarized and unpolarized light, we perform quantitative measurements of the force vector at each contact using an optimization technique \cite{abed_zadeh_enlightening_2019,kollmer_photo-elastic_nodate,liu_spongelike_2021,daniels_photoelastic_2017, fazelpour_effect_2021}. Via coarse-graining \cite{weinhart2013coarse, fazelpour_effect_2021}, we calculate the local shear stress $\tau(\Delta r)$ and local pressure $P(\Delta r)$ from these measured vector contact forces. We choose a width parameter $w=1.3d$ in our coarse-graining calculations to avoid both large fluctuations and over-smoothing (see \citet{fazelpour_effect_2021} for more details). The reported stress fields, $\tau(\Delta r)$ and $P(\Delta r)$, in Fig.~\ref{fig:stress} are time-averaged over $2000$ frames (taken over $\sim3$ hours), where vector contact forces and coarse-grained $\tau(\Delta r)$ and $P(\Delta r)$ are obtained for each frame.

Shear stress $\tau(\Delta r)$ and pressure $P(\Delta r)$ profiles are shown in Fig.~\ref{fig:stress} for all 6 wall shapes. The stress profiles are not reported near the inner and outer walls due to both imperfect lighting near the inner wall, as well as the $1.3 d$ the cutoff due to the coarse-graining length scale. As we can see in the stress profiles, even though we are using the same particles in all datasets and they rest on the same bottom plate, the stress profiles have different slopes depending on the choice of boundary conditions. This is most strongly apparent in the pressure profiles $P(\Delta r)$, for which we observe a slight slope instead of being constant, likely due to a combination of basal friction and humidity (which provide a weak adhesion between the particles and the bottom plate). Therefore, we find that it is important for all datasets to be collected at similar humidity (we selected a subset of runs which all have $\sim40\%$ humidity) in order to make direct run-to-run comparisons. For some wall shapes, we observe a slight increase in pressure (see Fig.~\ref{fig:stress}b) near the outer wall; this indicates that pressure can be applied on the particles due to wall roughness/compliance. 

\section{Results}

\subsection{Cooperative model validation}

As expected, we have observed that the choice of wall properties strongly controls the resulting flow (most apparent in the speed $v(\Delta r)$ and pressure $P(\Delta r)$ profiles), particularly near the outer wall. This sensitivity likely arises because the creeping flow (away from the shearing surface) is the most sensitive to nonlocal effects. 

\begin{figure}
	\centering 
	\includegraphics[width=0.9\linewidth]{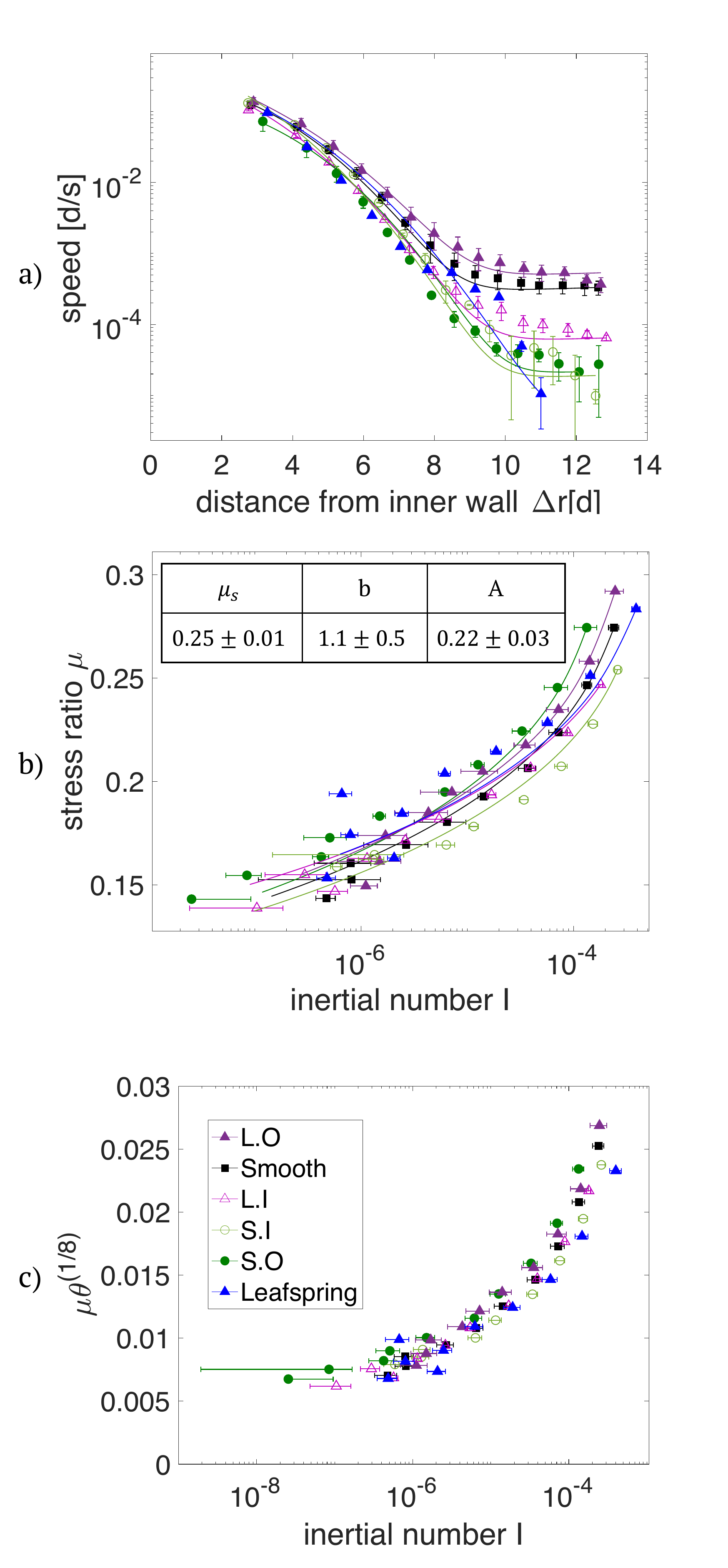} 
	\caption{(a) Speed profiles $v(\Delta r)$ as a function of distance from inner wall $\Delta r$, on a logarithmic scale. The data points represent experimentally measured flow speeds obtained from particle-tracking, for all 6 wall shapes. The error bars are the standard error of the particle speeds, and the solid lines are the cooperative model predictions. (b) Stress ratio $\mu$ as a function of inertial number $I$ for all 6 wall shapes. The data points are the experimentally measured $\mu$ from Eq.~\ref{eq:mu} and experimentally measured $I$ from Eq.~\ref{eq:I}. The error bars are the standard errors, and the solid lines are the cooperative model prediction of $\mu(I)$. Table inside (b) Nonlocal parameter $A$, local parameter $b$, and yield criterion $\mu_s$, with the $\pm$ values indicating the confidence interval for the parameters. These same fitting parameters are used in \citet{fazelpour_effect_2022}. (c) $\mu(I)$ data from (b) re-plotted, with $\mu$ rescaled by $\theta^{1/8}$ (nondimensionalized granular temperature) as predicted by the results of \citet{kim_power-law_2020}}
	\label{fig:model_comparison} 
\end{figure}

We first test the cooperative model to determine whether the same model and parameters can be used, independent of the wall shape/compliance. The model validation is done by comparing the model prediction with experimental results (see Fig.~\ref{fig:model_comparison}). With the speed $v(\Delta r)$ and shear rate ${\dot{\gamma}}(\Delta r)$ obtained from particle tracking and shear stress $\tau(\Delta r)$ and pressure $P(\Delta r)$ obtained from photoelastic measurements, we can calculate $ \mu(I)$ profiles using Eq.~\ref{eq:I} and Eq.~\ref{eq:mu}. The data points in Fig.~\ref{fig:model_comparison} represent experimentally-measured $v(\Delta r)$ and $\mu(I)$ for different walls. The cooperative model is shown as solid lines.
In the cooperative model, the fluidity $g$ is obtained by solving Eq.~\ref{eq:k3} using Matlab's ODE solver. The constitutive parameters $(A,b,\mu_{s})$ used in Eq.~\ref{eq:k3} match those used in \citet{fazelpour_effect_2022} for the same photoelastic particles, are given in Fig.~\ref{fig:model_comparison}, and are found to provide good fits, independent of wall roughness. Thus, we observe them to be well-founded material parameters.

The boundary conditions in Eq.~\ref{eq:k3} need to be determined experimentally, by knowing $g_{exp}$, as previously described in \citet{tang2018nonlocal}, \citet{fazelpour_effect_2022}. Once the fluidity $g(\Delta r)$ has been solved for, it can thereafter be used to calculate $\mu(I)$, from $I(\Delta r)=g(\Delta r)\mu(\Delta r)T$, and $v(\Delta r)$, from integrating $\dot{\gamma}=\mu(\Delta r)g(\Delta r)$. However, the boundary condition for solving $\dot{\gamma}=\mu(\Delta r)g(\Delta r)$ currently needs to be determined experimentally from measuring $v_{exp}$ near the outer wall. This is unfortunately not known from first principles.
This challenge is readily seen in Fig.~\ref{fig:speed}, where $v_{exp}$ near $R_o$ is very different for the six different wall shapes. Thus, the wall properties and measuring $v_{exp}$ near the outer wall are crucial for correctly implementing the cooperative model.

Fig.~\ref{fig:model_comparison}ab provides a comparison between the experimental results and cooperative model using a single set of parameters (Fig~\ref{fig:model_comparison}b). This comparison shows that the cooperative model agrees well with the data even in the very slow region of the flow (creeping region). This is apparent in both the speed profile $v(\Delta r)$ (Fig.~\ref{fig:model_comparison}a) and the rheology $\mu(I)$ (Fig.~\ref{fig:model_comparison}b). Therefore, the cooperative model can successfully capture the flow characteristics, and accommodate wall effects on the flow, if the wall slip has been independently measured. 
Thus, the key challenge remains that the cooperative model requires knowing wall slip to predict the flow, which depends on the wall properties. Currently, we have no first-principles method for predicting the wall slip. 

Last, it is important to note that we observed that measurements of the constitutive parameters $(A,b,\mu_{s})$ are sensitive to changes in humidity, and that it is advisable to collect particle-tracking and stress-field measurements under as similar conditions as possible.

\subsection{Force network fluctuations}

Understanding the effect of the wall properties on inter-particle interactions would advance our understanding of the controls on granular slip, and therefore also granular rheology. One mechanism by which this may be happening is the transmission of fluctuations via the force network. Our photoelastic particles provide a means to address how changes in internal stress propagation arise due to the wall properties. In \citet{thomas_force_2019}, we already observed that the force chains fluctuate in quasi-static regimes even where the particles are not rearranging. In this new investigation, we further test how the wall properties affect force chain fluctuations, focusing on regions near the outer wall (quasi-static regime). 

To measure the force fluctuations, we utilize light intensity $I$ as a proxy of force magnitude, rather than the fully-resolved vector forces. We split each image into concentric rings of width $0.57d$, and find the force intensity fluctuations by considering the intensity of all pixels within these rings $I_{\Delta r}$. We quantify the fluctuations within each ring $F({\Delta r})$ as the standard deviation of light intensity over both time and space:
\begin {equation}
F({\Delta r}) = \sqrt{\dfrac{\sum\limits_{t=1}^{T} (I_{{\Delta r},t}-\langle I_{{\Delta r},t}\rangle _{t,{\Delta r}})}{N_{\Delta r} T }}
\label{eq:FF}
\end {equation}
where the total number of frames is $T=2000$, $I_{{\Delta r},t}$ is the intensity of all pixels within each ring in $T$ frames, and $N_{\Delta r}$ is the number of pixels within each ring. To compare force fluctuation for various boundaries, we normalize the force fluctuation by average light intensity of the all particles over time $\frac{F({\Delta r})}{\langle I\rangle}$. This removes the effect of total pressure in the system.

\begin{figure} 
	\centering 
	\includegraphics[width=0.9\linewidth]{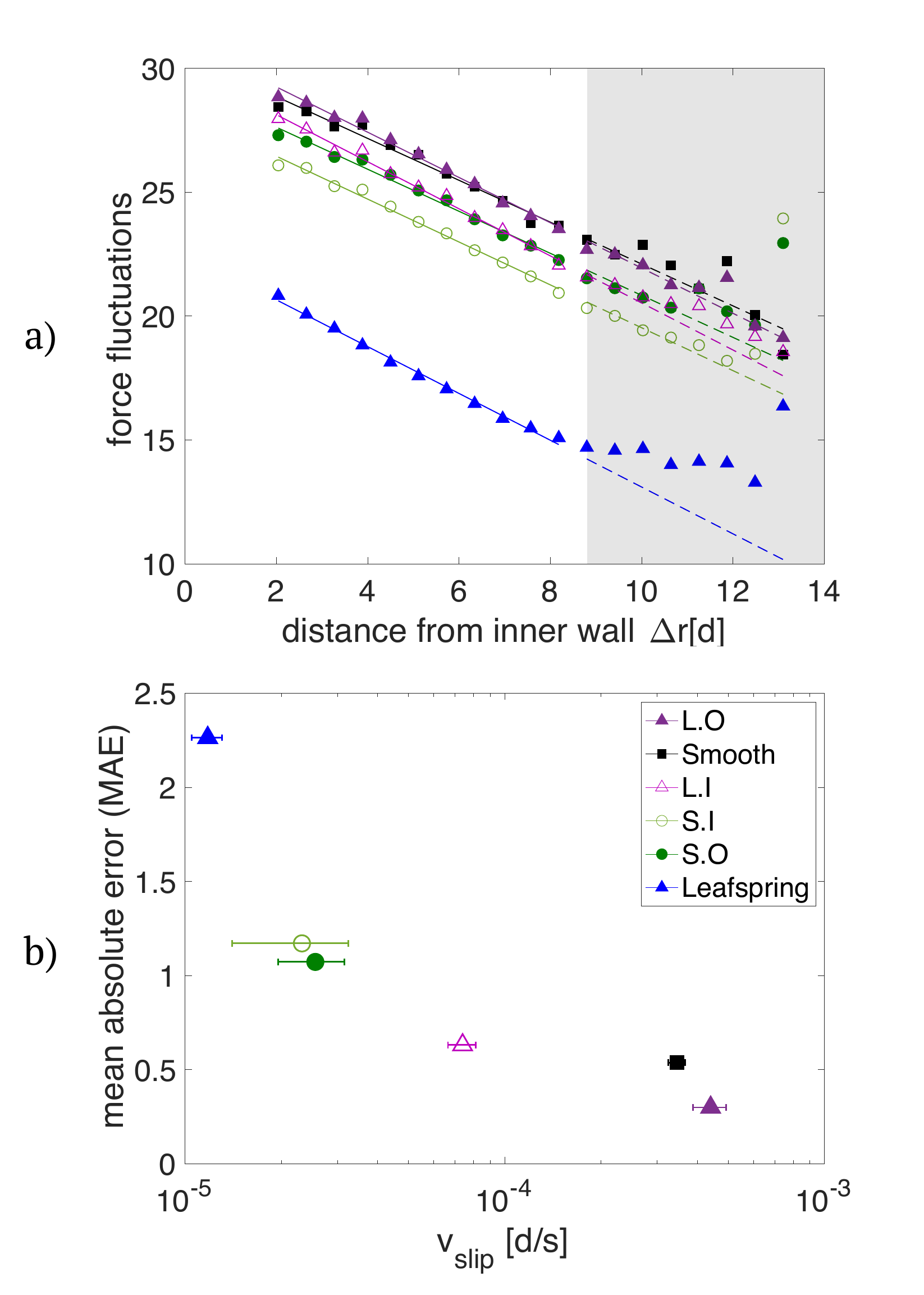} 
	\caption{(a) Normalized force fluctuations $F$ measured as a function of distance from inner wall for different wall shapes shown by data points. The solid lines indicate a linear fit to measured data in the white region. The gray shaded region correspond to the part of force fluctuation which deviates from linear trend. The dashed lines are the extension of original fitted line (solid lines) used to compare with measured data. (b) The mean absolute error (MAE) of force fluctuation near outer wall between measured data and fitted line on a $log$ scale.}
	\label{fig:force_fluc} 
\end{figure}

As shown in Fig.~\ref{fig:force_fluc}a, we observe that for all 6 wall shapes, the force fluctuations $F$ are understandably highest near the inner wall: the shearing wall generates force fluctuations as it moves. We further observe that the force fluctuations decrease approximately linearly while moving radially outward from the inner wall, and that this rate of decrease is similar for different wall roughness and compliance. However, the behavior near the outer (stationary) wall is quite different among the six runs, depending systematically on its roughness and compliance. 

We observe two main behaviors: for walls with larger wall slip (smooth, L.O., and L.I.), the force fluctuations decrease all the way to the outer wall at their approximately constant rate. However, for boundaries which have less slip at the wall (S.I., S.O., and leafspring), we instead observe a sharp increase in the force fluctuations starting a few particle diameters away from the outer wall. We interpret these fluctuations as arising from the roughness and/or compliance of the outer wall: the same feature that suppresses slip. Within this group of three, note that the two smoothest of these (smooth, L.O.) exhibit spatial oscillations away from the outer wall. This is likely associated with the formation of patches of ordered packings near the outer wall. These observations elucidate that force fluctuations can control granular slip, and could point toward a source of granular fluidity $g$ as a possible additional term to include in the ODE. 

To quantify this observed effect of the wall properties, we directly connect these excess force fluctuation to the measured wall slip $v_\mathrm{slip}$. First, we consider the inner region where force fluctuations drop at constant rate for all walls (white background in Fig.~\ref{fig:force_fluc}a) and make a linear fit $F'(\Delta r)$ (solid lines in Fig.~\ref{fig:force_fluc}a). Next, we extrapolate this fit into the outer region
where the force fluctuations are affected by the wall properties (gray background in Fig.~\ref{fig:force_fluc}a), shown as extensions dashed lines in Fig.~\ref{fig:force_fluc}a. The excess fluctuations above each dashed line can be quantified by calculating the mean absolute error (MAE) between the points and this line within the gray area:
\begin {equation}
\mathrm{MAE} = \langle{|F'(\Delta r)-F(\Delta r)|}\rangle
\label{eq:MAE}
\end {equation}
where $\langle \cdot \rangle$ is the average in the gray-shaded area in Fig.~\ref{fig:force_fluc}a. Larger MAE corresponds to observed force fluctuations deviations which are more in excess of the linear trend. As shown in Fig.~\ref{fig:force_fluc}b, we compare this value to the measured wall slip velocity at the outer wall, with $v_\mathrm{slip}$ taken as the average value of 3 data points ($\sim 1d$) closest to the outer wall, from Fig.~\ref{fig:speed}a. We observe that this plot exhibits a logarithmic-like relationship between the two values, with larger excess fluctuations corresponding to smaller wall slip. As a physical mechanism, it seems that particles can respond to external loads either by compressing, or by moving; the two ends of the graph differ in which one is the predominant response. Note that the data point for the smooth boundary appears somewhat out of order, as it is not the rightmost point on the graph. We believe this is due to the formation of patches of ordered packings that were mentioned above. While this plot allows us to empirically predict wall slip by knowing the force fluctuations, it is not yet directly tied to a quantified roughness.

\subsection{Rescaling $\mu(I)$ to a master curve}

While the $\mu(I)$ relation has been widely used to describe granular flows, the shapes of these curves vary extensively with varying system properties, particularly in quasi-static regions. Recently, \citet{kim_power-law_2020} proposed a new form for the $\mu(I)$ relationship, whereby flow profiles for various geometries collapse onto a master curve, and tested its validity for numerical simulations in 2D and 3D. 
They found that the shear stress ratio $\mu$ should be rescaled by the nondimensionalized granular temperature $\theta$, raised to some power. Motivated by the Chapman-Enskog relations, they examined a quantity $\theta \equiv \frac{\rho \, \delta v^2}{ D P}$ where $D$ is the number of spatial dimensions and $\delta v^2$ is the variance of the particle-scale velocity measurements (due to kinetic-theory-like fluctuations). We tested this prediction and found, in agreement with their results for 2D numerical simulations, that $\theta^{1/8}$ provides a good collapse to a universal curve, regardless of wall roughness. This data is shown in Fig.~\ref{fig:model_comparison}c. There is currently no theoretical prediction for the observed exponent $1/8$, but our experimental observations are consistent with the value obtained in their numerical simulations.

\section{Conclusions}
In this paper, we report on how dense granular flows can be controlled with boundary roughness and compliance. We test 6 different boundaries, 5 of which are rigid with different shapes and curvature, and one is deformable to test the effect of compliance. Using particle tracking, we observe that the speed profile near the inner wall is similar for all boundaries, while differing significantly near the outer wall. On the other hand, shear rate is not as strongly affected by boundary properties. Stress measurements are made throughout the material using photoelasticimetry which enables us to measure the full $\mu(I)$ profile. 

To test the efficacy of the cooperative model and its response to boundary roughness, we utilize the model parameters $(A,b,\mu_{s})$ previously optimized for our photoelastic particles. By comparing the model prediction to our measurements, we find that the cooperative model can capture the full flow profile and the effects of boundary properties on the flow. However, although the cooperative model is successful in describing granular flows, it is still necessary to {\it a priori} know the granular slip near the wall, in order to be able to predict flow.

We addressed this challenge by examining the inter-particle interactions arising through force chains. We find that dynamics of the force chains are strongly affected by the boundary properties, as the force chains fluctuate more near the rough boundaries than near the smooth boundaries. These effects are observed up to 5 particle diameter away from the wall. We establish a logarithmic relationship between excess force fluctuations and wall slip, providing an empirical tool for making qualitative predictions.

Moreover, we find, as observed in simulations of various geometries\cite{kim_power-law_2020}, rescaling $\mu(I)$ by a power of the nondimensionalized granular temperature provides a good collapse to a universal curve, regardless of boundary properties. This brings us one step closer to a general characterization of granular flows independent of geometry properties. In that regard, a next step is to test the cooperative model in various geometries, make flow measurements for a set of particles in one geometry and find the model parameters and then use them to provide predictions for flows in other geometries.

\section*{Conflicts of interest}
There are no conflicts to declare.

\section*{Acknowledgements}
We are grateful to the National Science Foundation (NSF DMR-1206808) for the construction of the particles and apparatus, and the International Fine Particle Research Institute (IFPRI) for financial support.
\balance
%\footnotesize{
%\bibliography{rheology_intro} %your .bib file
%\bibliographystyle{woc} %the RSC's .bst file
%}
%\bibliographystyle{secsty} %the RSC's .bst file
%\setcitestyle{square,aysep={},yysep={;}}
%\bibliography{rheology.bib}
%}

%\bibliography{r.bib}

\providecommand*{\mcitethebibliography}{\thebibliography}
\csname @ifundefined\endcsname{endmcitethebibliography}
{\let\endmcitethebibliography\endthebibliography}{}

%\bibliographystyle{woc} %the RSC's .bst file
%\end{itemize}

\end{document}